\documentclass[%
 reprint,
showpacs,preprintnumbers,
 amsmath,amssymb,
 aps,
prb,
]{revtex4-1}

\usepackage{graphicx}
\usepackage{dcolumn}
\usepackage{color}      
\usepackage{bm}
\usepackage{multirow}

\def \ang{\,\rm\mathring{A}}
\def \bcs{$\rm BaCuSi_{2}O_{6}$}
\def \cso{$\rm Cu_{2}(SiO_{3})_{4}$}
\def \cucu{$\rm Cu-Cu$}
\def \cuo4{$\rm CuO_{4}$}

\begin{document}

\preprint{APS/123-QED}

\title{Two types of adjacent dimer layers in the low temperature phase of $\mathbf {BaCuSi_{2}O_{6}}$}

\author{D.~V.~Sheptyakov}
\email{Denis.Cheptiakov@psi.ch}

\author{V.~Yu.~Pomjakushin}
\affiliation{ Laboratory for Neutron Scattering, Paul Scherrer
Institut, CH-5232 Villigen PSI, Switzerland
}%

\author{R.~Stern}
\email{Raivo.Stern@kbfi.ee}

\author{I.~Heinmaa}
\affiliation{ National Institute of Chemical Physics and Biophysics,
Tallinn 12618, Estonia
}%

\author{H.~Nakamura}

\author{T.~Kimura}
\affiliation{ Division of Materials Physics, Graduate School of
Engineering Science, Osaka University, Toyonaka, Osaka 560-8531,
Japan
}%

\date{\today}

\begin{abstract}
The low-temperature crystal structure of $\rm BaCuSi_{2}O_{6}$\ has
been investigated with high-resolution synchrotron x-ray and neutron
powder diffraction techniques and has been found to be on average
(ignoring the incommensurate modulation) orthorhombic, with the most
probable space group $Ibam$. The $\rm Cu-Cu$\ dimers in this
material are forming two types of layers with distinctly different
interatomic distances. Subtle changes also modify the partially
frustrated interlayer $\rm Cu-Cu$\ exchange paths. The present
results corroborate the interpretation of low-temperature nuclear
magnetic resonance and inelastic neutron scattering data in terms of
distinct dimer layers.
\end{abstract}

\pacs{61.05.C-, 61.05.fm, 61.66.Fn, 64.60.-i}


\maketitle


\section{\label{Intro}INTRODUCTION}

Intense studies of field--induced quantum phase transitions (QPTs)
in various magnetic insulators continue to enrich our knowledge of
the possible quantum ground states of matter.~\cite{Affleck91,
Giamarchi99, Sachdev99, Rice02} Structurally dimerized quantum spin
systems play a leading role in these studies; the classes of
field--induced QPTs known to date in these systems are well
summarized in Refs.~\onlinecite{RueggINS2007,GRT-2008}. For magnetic
interactions with weak or no frustration, the kinetic energy of the
triplet quasiparticles in such systems is dominant and the ordered,
or Bose-Einstein condensed (BEC), ground state is uniform at $H >
H_c$, as e.g. in TlCuCl$_3$.~\cite{Rice02, Ruegg-Nature-2003}
In SrCu$_2$(BO$_3$)$_2$~\cite{Rice02, Kageyama99,
Kodama-Science-2002} triplet hopping is suppressed by geometrical
frustration and the repulsion causes the condensed triplets to form
a superlattice with the appearance of magnetization plateaus. To
better distinguish between the various scenarios and to determine
accurately the relevant exchange couplings adequate knowledge of
crystal structures and symmetries is of outmost relevance.

\begin{figure}
\includegraphics[width=5cm]{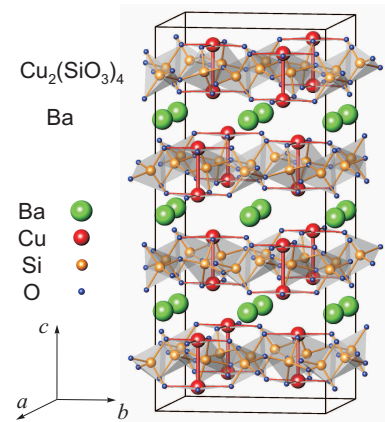}
\caption{\label{fig_01}(Color online) General view of the room
temperature tetragonal crystal structure of \bcs. Copper silicate
\cso\ layers are interleaved with the layers of Ba atoms. The
drawing is based on the crystal structure parameters refined from
neutron powder diffraction data collected at $T=130\,K$.}
\end{figure}

The interest in \bcs\ is motivated by its extraordinary phase
diagram with field-induced
BEC.~\cite{Jaime-prl-2004,Sebastian-Nature-2006,GRT-2008} Being a
quantum paramagnet at zero magnetic field down to the lowest
temperatures, the system displays a QPT into a magnetically ordered
state at the critical value of magnetic field of
$\rm\sim23.5\,T$.~\cite{Jaime-prl-2004,Sebastian-Nature-2006} At
$\rm\sim610\,K$, the crystal structure of \bcs\ undergoes a phase
transition from the high-temperature (HT) phase with space group
$I4/mmm$ and lattice parameters $a_{ht}\sim7.11\ang$,
$c_{ht}\sim11.2\ang$ to its room-temperature (RT) phase with a space
group $I4_1/acd$ and lattice parameters
$a_{rt}\sim\sqrt{2}*a_{ht}\sim10\ang$, $c_{rt}\sim
2*c_{ht}\sim22.5\ang$.~\cite{Sparta-2004} In the essentially layered
RT tetragonal crystal structure of
\bcs,~\cite{Finger-1989,Sparta-2004} the copper silicate \cso\
layers are separated by the intermediate layers of $\rm Ba$ atoms
(illustration in Fig.~\ref{fig_01}). The \cucu\ dimers with an
interatomic distance of $\sim2.75\ang$ are well separated from each
other in the structure, since the inter-dimer \cucu\ distances
within the layer ($\sim7\ang$) and the \cucu\ distances between the
dimers in adjacent layers ($\sim5.75\ang$) are much larger. Thus, to
a great extent, the ground state of the system is determined by the
\cucu\ interactions within the
dimers.~\cite{Sasago-prb-1997,Zvyagin-prb-2006}

At zero magnetic field, the ($S=1/2$) $\rm Cu^{2+}$ ions are paired
antiferromagnetically, while the degenerate excited triplet states
with $S=1$ are gapped;~\cite{Sasago-prb-1997} thus the compound
possesses an essentially singlet ground state with $S=0$ at $H=0$.
Upon application of a magnetic field, the excited triplet states are
Zeeman split with the energy scale of the splitting being
proportional to magnetic field. At a critical magnetic field value
of $H_{c}\sim23.5\,\rm T$, the energy of the lowest triplet state
with $S_z=+1$ becomes lower than that of a singlet state with $S=0$,
and a magnetic field-induced BEC of the excitations occurs. The
population of the bosons (measured as the magnetization of the
sample) may be precisely tuned by the magnetic field. Another
intriguing phenomenon observed in \bcs\ next to this QPT is
dimensional crossover around 1~K from a 3D into a 2D regime with
lowering temperature.~\cite{Sebastian-Nature-2006, Batista-prl-2007}
To explain this feature, perfect frustration between adjacent \cso\
layers was assumed.~\cite{Sebastian-Nature-2006, Batista-prl-2007,
SchmalBat-prb-2008} Alternatively, the hypothesis has emerged that a
structural modulation along the $c$ axis~\cite{Horvatic-ptp-2005,
RueggINS2007, Kraemer-PRB-2007} combined with nonperfect frustration
is the sole reason for the lowered
dimensionality.~\cite{RoschVoita_prb7618, RoschVoita_prb7622} As a
unifying theory, a possibility has been
proposed~\cite{Laflorencie-2009} that the combined effect of the
structural modulation along the $c$ axis and the nearly perfect
interlayer frustration is a reason for the observed dimensional
crossover.

In attempts to build a theory of this QPT, the compound was
originally assumed to possess the same square motif of the \cucu\
dimers at low temperatures as at room
temperature~\cite{Sebastian-prb-2005, Sebastian-Nature-2006,
Jaime-prl-2004} resulting in three relevant coupling constants
entering the spin Hamiltonian: the intradimer, interdimer, and
interlayer couplings.

However, high resolution inelastic neutron scattering (INS) data
were clearly at odds with this picture.\cite{RueggINS2007}
Furthermore, a first-order structural phase transition at
$\rm\sim100\,K$ has been discovered by Stern\hspace{3pt}\it
et\hspace{3pt}al.\rm~\cite{Kraemer-PRB-2007} and reported by
Samulon\hspace{3pt}\it et\hspace{3pt}al.\rm~\cite{Samulon-prb-2006}
which necessarily states that the true spin Hamiltonian of the
system is more complex than originally thought. The method of single
crystal diffraction used in Ref.~\onlinecite{Samulon-prb-2006}
confirmed the transition into an orthorhombic (or weakly monoclinic)
structure. Moreover, it allowed for observation of satellite peaks
indicating an incommensurate modulation in the low-temperature (LT)
phase of \bcs. Here we use in contrast a combination of
high-resolution x-ray and neutron \emph{powder} diffraction
techniques in order to determine the \emph{average} crystal
structure of the LT phase of \bcs.

\section{\label{Expe}EXPERIMENTAL}

\subsection{\label{Synthe}Synthesis}

Polycrystalline samples of \bcs\ were prepared by solid-state
reaction. Powders of  $\rm BaCO_{3}$, $\rm CuO$, and $\rm SiO_{2}$
were weighted to the prescribed ratios, mixed, and well ground. The
mixture was calcined at 900$^\circ$C in air for 20~h. The resulting
powders were pulverized, isostatically pressed into a rod
shape ($\sim5$~mm diameter, $\sim50$~mm length) and sintered again
at 1010$^\circ$C in air for 20~h.

\subsection{\label{Powdiffresults}Powder diffraction experiments}

The powder diffraction experiments have been carried out with two
high-resolution diffractometers: the Powder Diffraction station of
the Materials Sciences Beamline (MS-PD) at the Swiss Light
Source,~\cite{sls-MS-PD} and the high-resolution powder neutron
diffractometer HRPT~\cite{Fischer00} at the spallation neutron
source SINQ, both at Paul Scherrer Institute in Villigen.

At the MS-PD station, the synchrotron x-ray diffraction data were
collected on a powder sample enclosed in a capillary of a $0.3$~mm
diameter, which was placed in a Janis flow-type cryostat. The
Microstrip Mythen-II detector was used, which allowed for high
counting rates while maintaining the high resolution which was
essentially sample-conditioned. The typical counts of $\sim2*10^{5}$
in the strongest peaks were achieved within $\sim1$~minute. On
cooling, the diffraction patterns were collected with 1~K steps in
the temperature range from 154 to 64~K, and with 2~K steps in the
temperature range from 64 to 12~K. After collecting the lowest
temperature dataset at 4~K, the data were collected on heating the
sample in the temperature range from 5 to 123~K with typically
$1\dots2$~K steps and from 128 to 178~K with 5~K steps, in order to
observe the transition both on cooling and on heating and to
characterize the hysteresis thereof.

At HRPT, a bigger amount ($\sim 1\,g$) of the very same powder
sample of \bcs\ was enclosed into a vanadium can of a 6~mm diameter,
and the data were collected using a close-cycle refrigerator at
130~K (well above the transition on cooling) and at 13~K (far below
the transition) in the instrument setup with the highest resolution.
The ultimate resolution was needed because the orthorhombic
splitting of the peaks of the parent tetragonal RT structure
characterized by an orthorhombicity parameter defined as
${2*|a_{orth}-b_{orth}|}/(a_{orth}+b_{orth})$ amounts only to
$\sim1.7*10^{-3}$.

The data of the neutron diffraction experiment were used for the
structure determination, and for the precise refinement of the
structural parameters at low temperature, while the synchrotron
x-ray powder diffraction data have mainly been used for indexing,
space group selection and refinement of the temperature dependence
of the main structural parameters, as well as for identifying all
the impurity phases present. The following impurities have been
determined in the powder sample: $\rm BaCu_{2}Si_{2}O_{7}$
($\sim4.4~\%~wt.$), $\rm BaCuSi_{4}O_{10}$ ($\sim1.3~\%~wt.$), $\rm
Ba_{4}Si_{6}O_{16}$ ($\sim1.1~\%~wt.$), $\rm BaSiO_3$
($\sim0.3~\%~wt.$), and $\rm Cu_2O$ ($\sim0.1~\%~wt.$).
Approximately $93~\%~wt.$ of the powder sample is the main \bcs\
phase. While treating the neutron data with $\sim5000$ counts in the
strongest peak, the only relevant impurity which was definitely seen
in the diffraction pattern and correspondingly input into the
refinements was $\rm BaCu_{2}Si_{2}O_{7}$ ($\sim4.4~\%~wt.$); the
others were undetectable, and thus were not considered.

The RT crystal structure model (Ref.~\onlinecite{Sparta-2004}) has
been unambiguously confirmed by the Rietveld refinements based on
both synchrotron x-ray and neutron data above the phase transition
temperature. We do not focus on this issue here, but rather
concentrate on the LT structure.

\section{Results}

\subsection{\label{LT-str}Low-temperature crystal structure of $\mathbf {BaCuSi_{2}O_{6}}$}

\begin{figure}
\includegraphics[width=4.25cm]{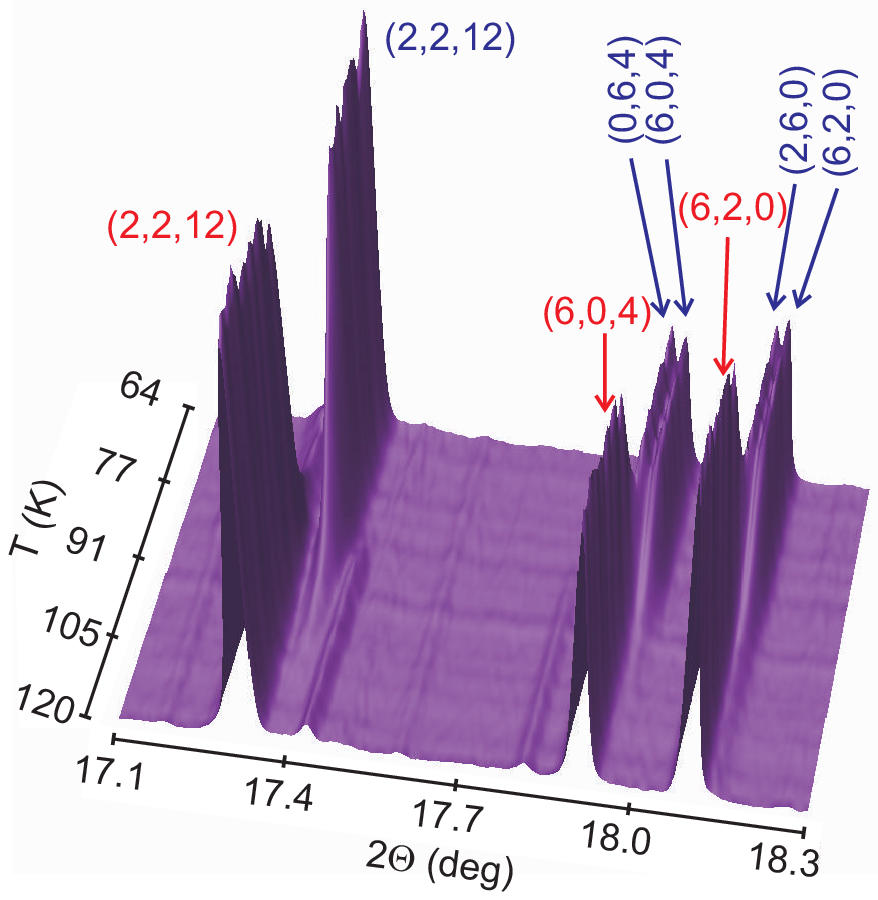}
\includegraphics[width=4.25cm]{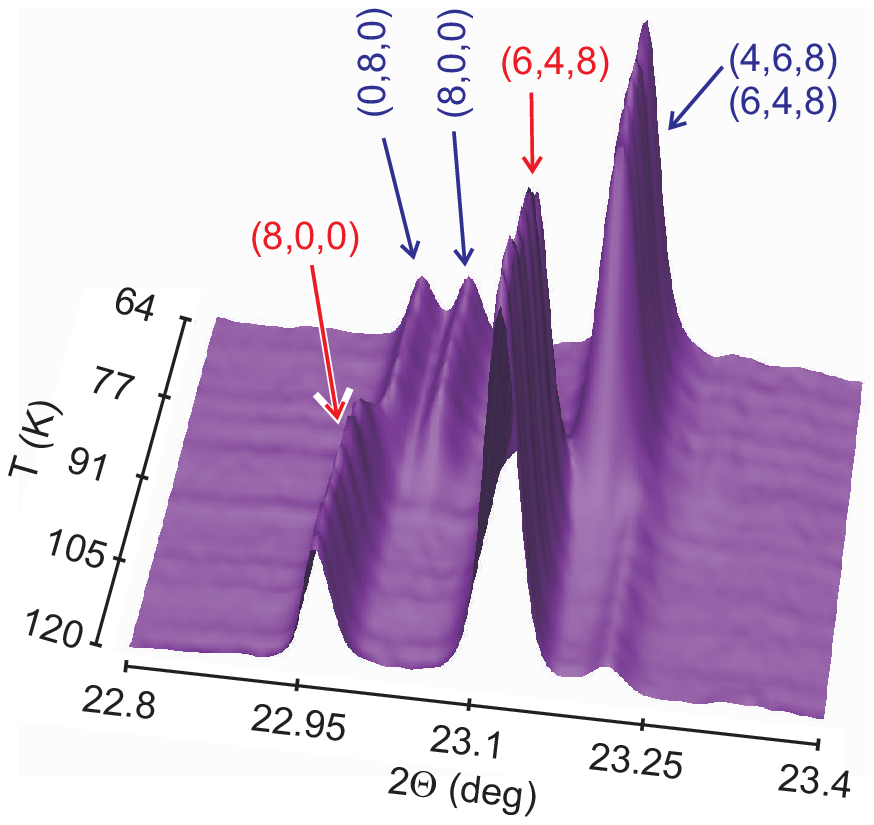}
\caption{\label{fig_02}(Color online) Selected ranges of the
synchrotron x-ray powder diffraction patterns of \bcs\ taken with
1~K steps on cooling from 120 to 64~K. Obvious are the jump in the
unit cell volume at the transition at $\sim89.7$~K (all the main
peaks are shifted to higher $2\Theta$ angles) and the splitting of
the diffraction peaks with $h\neq k$.}
\end{figure}

\begin{figure}
\includegraphics[width=5cm]{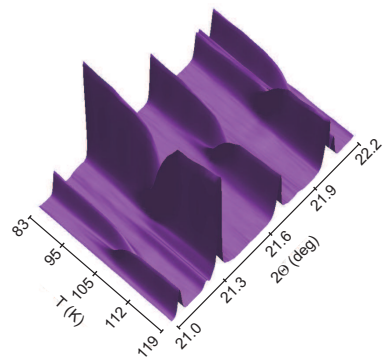}
\caption{\label{fig_03}(Color online) Selected range of the
synchrotron x-ray powder diffraction patterns of \bcs\ taken with
2~K steps on heating from 83 to 97~K and with 1~K steps on heating
from 99 to 119~K (note the nonuniform scale on the temperature
axis). One clearly sees the structural phase transition occurring
(on heating) at $\sim107.2$~K.}
\end{figure}

The LT structural phase transition in \bcs\ is clearly seen in both
neutron and x-ray diffraction patterns, and is most evident in the
temperature-dependent x-ray data. Illustrations of the transition
are given in Figs.~\ref{fig_02} and~\ref{fig_03}. While all the
peaks get a little narrower when cooling below the transition, we do
not observe any of them to broaden. Instead, some remain single
peaks, and some are splitting. The peak $(2,2,12)$ at
$\sim17.35^\circ$ remains single, while the $(6,0,4)$ and $(6,2,0)$
at $\sim17.9^\circ$ and $\sim18.15^\circ$ become split into the
orthorhombic pairs $(0,6,4)/(6,0,4)$ and $(2,6,0)/(6,2,0)$,
correspondingly. The peak $(8,0,0)$ at $\sim23^\circ$ splits into a
$(0,8,0)/(8,0,0)$ doublet, and the strong $(6,4,8)$ at
$\sim23.1^\circ$ transforms into a pair of $(4,6,8)/(6,4,8)$ which
are not resolved, since the amplitudes of their scattering vectors
are too close to each other. The backward phase transition on
heating manifests itself (see Fig.~\ref{fig_03}) as a recovery of
the diffraction pattern to that of the RT crystal structure, yet it
occurs at a higher temperature.

The LT synchrotron x-ray diffraction pattern of \bcs\ showing clear
splitting of the diffraction peaks $(hkl)$ with $h\neq k$ of the
parent room temperature tetragonal crystal structure (space group
$I4_1/acd$) has been indexed on the orthorhombic unit cell with
lattice constants (at $T=4\,\rm K$) $\{a,b,c\}=\{9.951\ang,
9.968\ang, 22.239\ang\}$. From the systematic absences it has been
found that the most probable space group is body-centered, and all
the relevant possibilities have been checked for the structure
determination. In fact we have performed the structure determination
for \emph{all} space groups and combinations of $(a,b,c)$, which did
account for all diffraction peaks observed, although for many of
these some of the calculated peaks were actually absent in the
experimental pattern. Having done so, we may exclude any chance for
having missed the true solution. The structure determinations have
been carried out on a neutron diffraction dataset taken at
$T=13\,\rm K$, either with the program FOX~\cite{Fox} or by direct
symmetry reduction in cases where it was possible and
straightforward. All the models have been checked with extensive
Rietveld refinements (done with the program {\tt
FULLPROF}~\cite{Fullprof93}) of their parameters from both the
synchrotron x-ray and neutron powder diffraction patterns.

\begin{figure}
\includegraphics[width=\columnwidth]{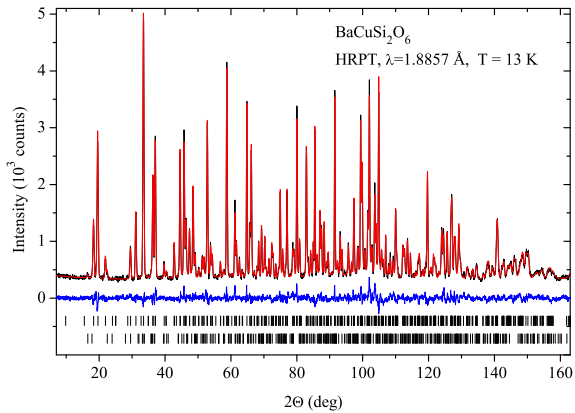}
\caption{\label{fig_04}(Color online) Rietveld refinement of the
crystal structure parameters of the low-temperature (LT) phase of
\bcs\ from the neutron powder diffraction data. The observed
intensity, calculated profile, and difference curve are shown. The
rows of ticks at the bottom correspond to the calculated diffraction
peak positions of the phases (from top to bottom): \bcs\ -
orthorhombic LT structure, $\rm BaCu_{2}Si_{2}O_{7}$ impurity
($\sim4.4~\%~wt.$)}
\end{figure}

\begin{figure}
\includegraphics[height=5.7cm]{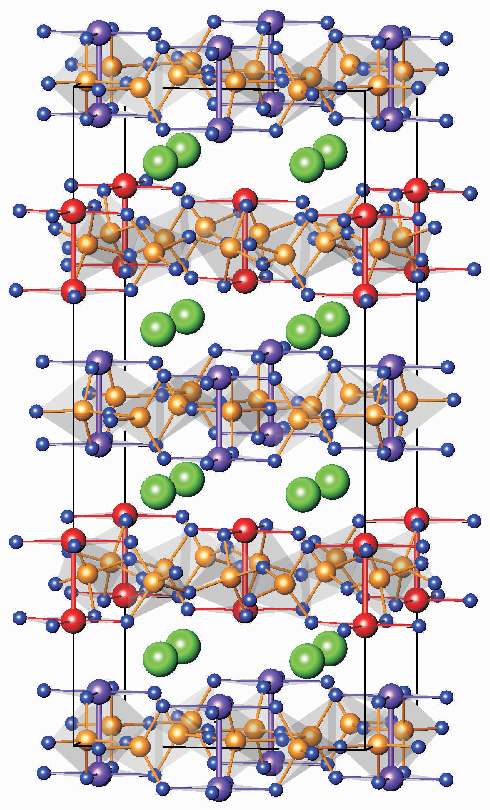}
\includegraphics[height=5.7cm]{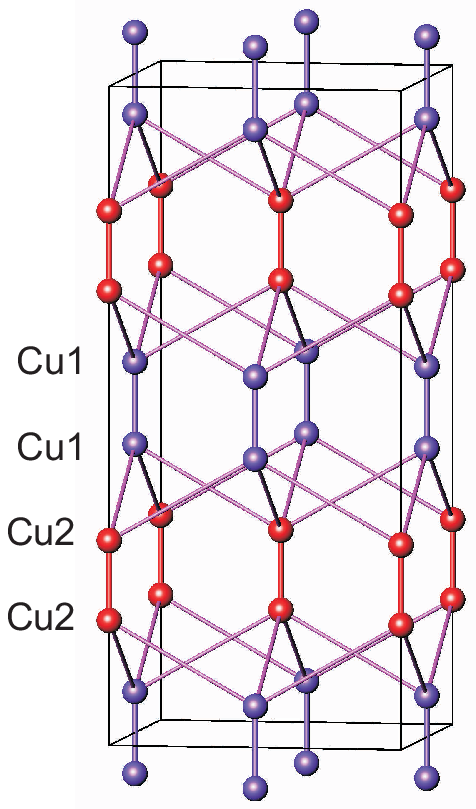}
\caption{\label{fig_05}(Color online) Schematic representation of
the crystal structure of the LT phase of \bcs. In the right panel,
only the \cucu\ dimers are shown. The $\rm Cu1-Cu1$ distance of
$2.774(6)\ang$ is much longer than the $\rm Cu2-Cu2$ with
$2.701(6)\ang$. For the inter-layer coupling paths, $\rm Cu1-Cu2$
(shown in thin lines in the right figure), there are always two
pairs of slightly inequivalent bond distances (see
Table~\ref{tab:table_2}).}
\end{figure}

While going from higher to lower symmetries in the orthorhombic
syngony, already the space group $Ibca$ (No.~73), which is a direct
subgroup of the RT structure space group $I4_1/acd$, gives the first
sensible solution. Yet the next group $Ibam$ (No.~72) supplies a
definitely better model. Compared to the $Ibca$ model, the model in
$Ibam$ provides a significant improvement of the agreement factors:
$\chi^2$ decreases from 4.33 to 2.66, and the Bragg R-factor -- from
7.05 to 4.80. Both structure models are in fact slight distortions
of the RT tetragonal structure with $I4_1/acd$ symmetry, but are
having very essential difference between each other from the point
of view of the geometry of the \cucu\ dimer layers and their
stacking along the direction perpendicular to the \cso\ layers. The
unique Cu atom in the RT structure model is being split into two
distinct atoms in both models, but in the case of the $Ibca$ model
they are forming just one type of \cucu\ dimer layer, while the
correct $Ibam$ model contains two distinctly different \cucu\ dimer
layers, and these will be discussed below. The Rietveld refinement
plot obtained with the neutron data is shown in Fig.~\ref{fig_04}.
The refined parameters of the LT crystal structure are given in
Table~\ref{tab:table_1}, and the most relevant interatomic distances
in Table~\ref{tab:table_2}. The schematic representation of the
crystal structure of the LT phase of \bcs\ is given in
Fig.~\ref{fig_05}.

\begin{table}
\caption{\label{tab:table_1} The refined crystal structure
parameters of the LT phase of \bcs. Results obtained from the
refinement done on neutron powder diffraction data collected at 13~K
with the wavelength $\lambda=1.8857\ang$. Space group $Ibam$
($\#72$, Z=16). Isotropic temperature factors were refined with a
constraint to equality for Cu atoms, for Si atoms, for the $\rm
O1\dots O4$ atoms (``apical'' ones of the $\rm SiO_4$ tetrahedra,
also coordinating Cu in the $\rm CuO_4$ squares), and for the $\rm
O5\dots O7$ atoms (directly bridging the Si atoms, i.e. located at
exactly or roughly the same $z$ values as Si). Atoms are located in
the following crystallographic positions: Cu1 -- in the
$8i\,(0,0.5,z)$, Cu2 -- in the $8h\,(0,0,z)$, Si1 -- in the
$8g\,(0,y,1/4)$, Si2 - in the $8f\,(x,0,1/4)$, Si3, Si4, O6 and O7 -
in the $8j\,(x,y,0)$ sites correspondingly; all the others - in the
general $16k$ positions $(x,y,z)$. All positional parameters were
refined without any constraints.}
\begin{ruledtabular}
\begin{tabular}{|l|c|c|c|c|}
\multicolumn{5}{|c|}{\bcs\ Lattice parameters at $\rm T=13\,K$}\\
\hline
$a,\ang$&\multicolumn{4}{c|}{9.95129 (7)}\\
$b,\ang$&\multicolumn{4}{c|}{9.96792 (7)}\\
$c,\ang$&\multicolumn{4}{c|}{22.23914 (13)}\\
$\rm V,\ang^3$&\multicolumn{4}{c|}{2205.98 (3)}\\
\hline
Atom&x&y&z&$B_{iso},\,\rm\mathring{A}^2$\\
\hline
Ba&0.2567 (5)&0.2444 (5)&0.1234 (3)&0.09 (5)\\
Cu1&0&0.5&0.0624 (2)&0.60 (3)\\
Cu2&0&0&0.1893 (2)&0.60 (3)\\
Si1&0&0.2761 (11)&0.25&0.29 (4)\\
Si2&0.2773 (12)&0&0.25&0.29 (4)\\
Si3&0.0019 (9)&0.7724 (11)&0&0.29 (4)\\
Si4&0.7771 (13)&0.0067 (9)&0&0.29 (4)\\
O1&0.0362 (5)&0.8095 (5)&0.3089 (3)&0.77 (3)\\
O2&0.3054 (6)&0.4622 (5)&0.1891 (3)&0.77 (3)\\
O3&0.3076 (5)&0.0092 (5)&0.0608 (3)&0.77 (3)\\
O4&0.5123 (5)&0.1918 (5)&0.0626 (4)&0.77 (3)\\
O5&0.1281 (6)&0.3730 (5)&0.2698 (2)&1.09 (4)\\
O6&0.3454 (7)&0.3454 (7)&0&1.09 (4)\\
O7&0.1042 (7)&0.1043 (7)&0&1.09 (4)\\
\end{tabular}
\end{ruledtabular}
\end{table}

\begin{table}
\caption{\label{tab:table_2} The most relevant interatomic distances
($\ang$) in the crystal structure of the LT phase of \bcs\ refined
from the neutron powder diffraction data at $\rm T=13\,K$.}
\begin{ruledtabular}
\begin{tabular}{|c|c|}
Cu1-Cu1&2.774 (6)\\
Cu2-Cu2&2.701 (6)\\
Cu1-Cu2&5.728 (3) and 5.720 (2)\\
$\rm<Si1-O(1,1,5,5)>$&1.632 (16)\\
$\rm<Si2-O(2,2,5,5)>$&1.634 (18)\\
$\rm<Si3-O(4,4,6,7)>$&1.63 (2)\\
$\rm<Si4-O(3,3,6,7)>$&1.61 (2)\\
\end{tabular}
\end{ruledtabular}
\end{table}

Unlike in the single crystal experiments reported in
Ref.~\onlinecite{Samulon-prb-2006}, our study carried out by powder
diffraction could not reveal the presence of any incommensurate
peaks due to their relative weakness: According to
Ref.~\onlinecite{Samulon-prb-2006}, the central Bragg peaks did have
to be scaled down by a factor of $\sim10^{4}$ to display on the same
scale with the satellite incommensurate modulation peaks. Such
dynamic ranges are inaccessible for powder diffraction, and in this
sense, the presence of the incommensurate modulation did not hinder
the determination of what we believe to be a most probable model for
the \emph{average} LT crystal structure of \bcs.

\begin{figure}
\includegraphics[height=4cm]{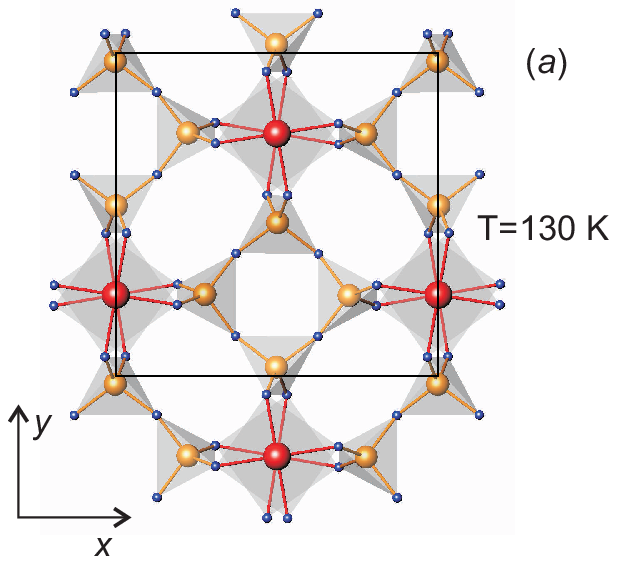}
\includegraphics[height=4.2cm]{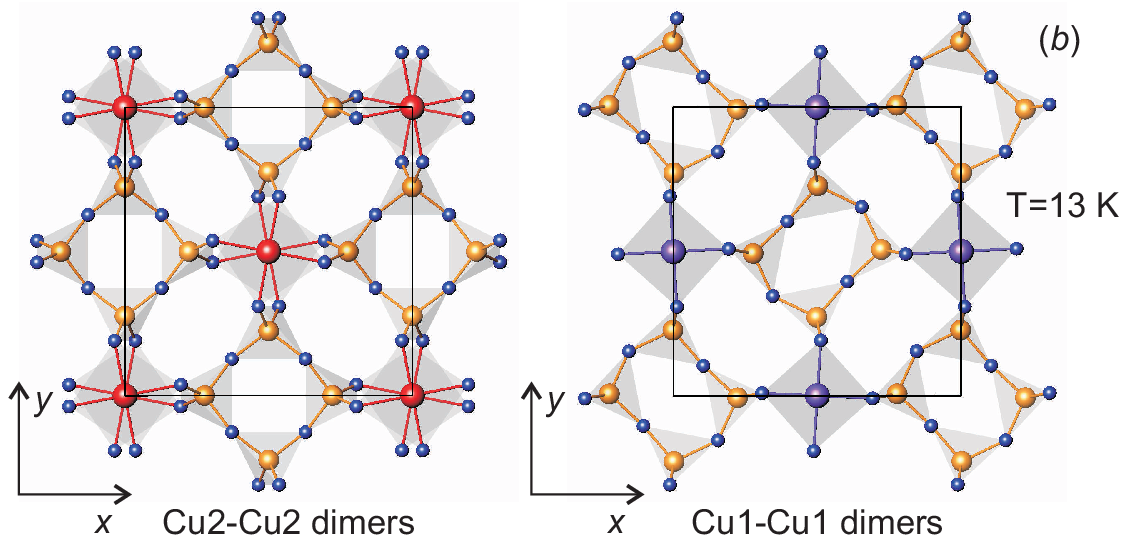}
\caption{\label{fig_06}(Color online) Cu-Cu dimer layer structure.
(\emph{a}): In tetragonal phase (the figure is based on the atom
positions refined from the neutron data at 130~K). \cucu\ distance
is $2.731(2)\ang$. (\emph{b}): Two types of \cucu\ dimer layers at
low temperature as refined from the neutron data taken at 13~K. The
$\rm Cu2-Cu2$ distance is $2.701(6)\ang$; the $\rm Cu1-Cu1$ distance
is $2.774(6)\ang$.}
\end{figure}

An interesting feature is observed in the geometry of the \cso\
layers which is very relevant for the interpretation of
low-temperature magnetic properties: Mutual rotation of the \cuo4\
coordination polyhedra in the Cu-dimer layers, characteristic for
the RT structure on \bcs, is only preserved in the layers with
shorter \cucu\ dumbbell distances. This feature is illustrated in
Fig.~\ref{fig_06}: At 130~K, in the tetragonal phase, all the \cucu\
dimer layers are equivalent and the \cuo4\ coordination polyhedra
look like squares slightly turned ($\sim19^\circ$) with respect to
each other in each \cucu\ dumbbell around its axis
[Fig.~\ref{fig_06}~(\emph{a}): view along the \emph{c} direction of
the unit cell]. At low temperature, only the $\rm Cu2-Cu2$ dimer
layer with shorter \cucu\ distances, shown on the left of
Fig.~\ref{fig_06}~(\emph{b}), preserves this feature (the mutual
rotation angles of the $\rm Cu2O_4$ squares with respect to each
other are on the order of $\sim21\ldots22^\circ$), and the $\rm
Si_{4}O_{12}$ rings in these layers are also (same as at 130~K)
nearly squares. In contrast to this, in the $\rm Cu1-Cu1$ dimer
layer with longer \cucu\ distances, shown on the right of
Fig.~\ref{fig_06}~(\emph{b}), the \cuo4\ coordination polyhedra are
not turned with respect to each other. The $\rm Si_{4}O_{12}$
4-member rings in the $\rm Cu1-Cu1$ dimer layers show clear
deviations from the ideal square shape, forming a checkerboard type
of ordering of their elongation directions along one of the $a-b$
plane diagonals. The intralayer \cucu\ distances are identical in
both types of layers in our model, yet it is too hard to quantify
the obvious differences in the exact intralayer exchange paths
between the dimers, which are obviously nonequivalent for the layers
with the shorter and the longer \cucu\ dimers.

\subsection{\label{trans}First-order character of the transition}

\begin{figure}
\includegraphics[width=\columnwidth]{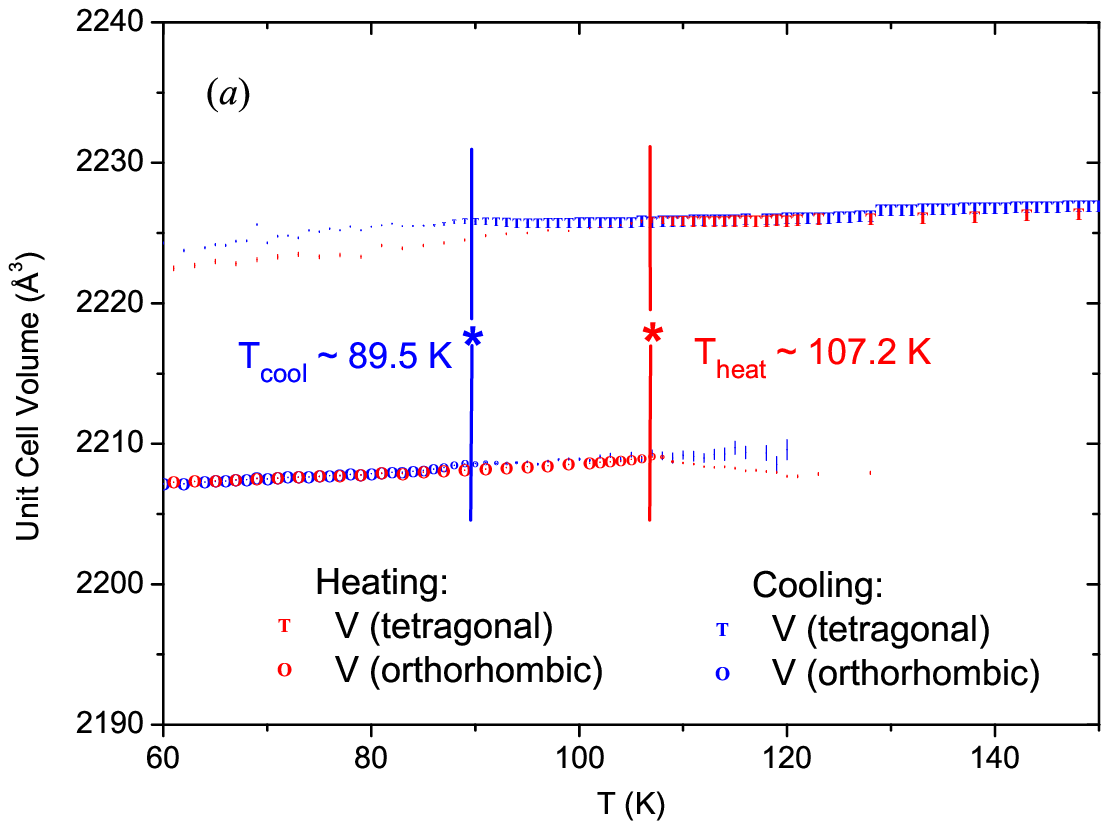}
\includegraphics[width=\columnwidth]{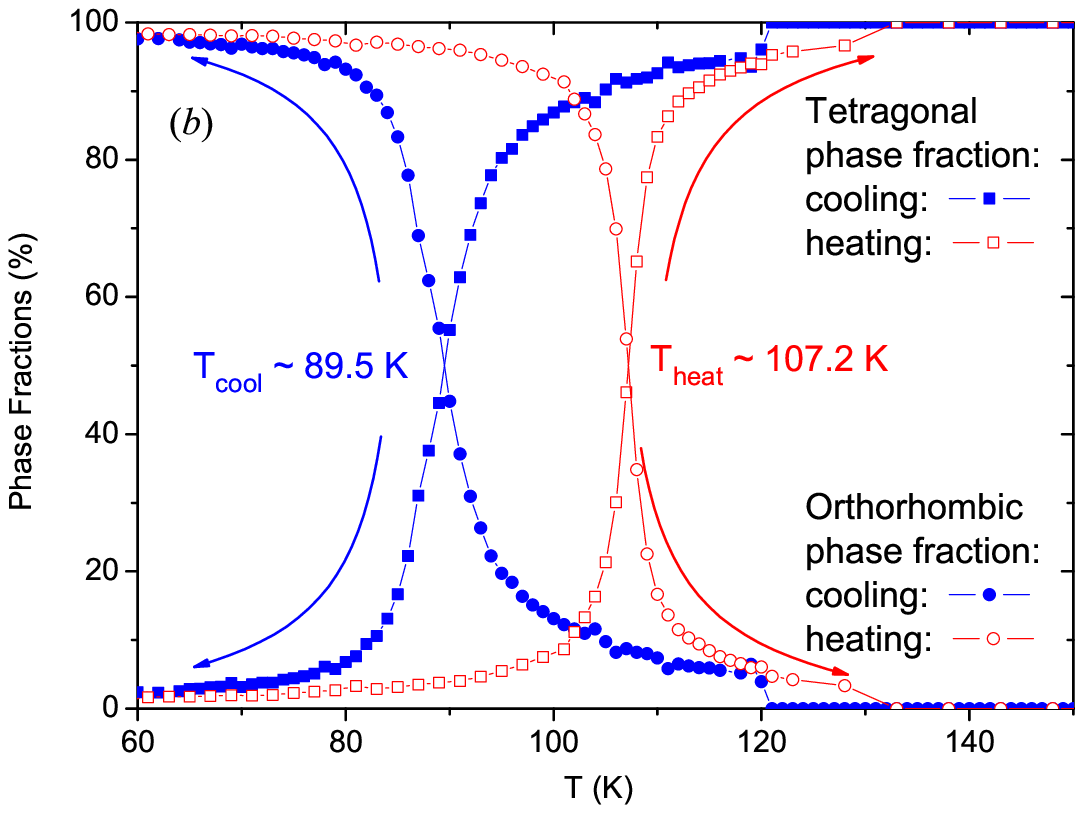}
\caption{\label{fig_07}(Color online) Unit cell volumes on cooling
and on heating (\emph{a}) and the relative phase weight proportions
(\emph{b}) for the tetragonal and orthorhombic modifications of
\bcs. Results of the refinements done with the synchrotron x-ray
data. Symbol sizes (T) and (O) in panel (\emph{a}) are proportional
to the corresponding weight phase fractions. Error bars in
(\emph{a}) are also shown and are typically much smaller than the
symbol sizes.}
\end{figure}

In agreement with the results already reported in
Ref.~\onlinecite{Samulon-prb-2006}, the structural transition we
observe in \bcs\ is of first order. The refined unit cell volumes,
as well as the weight proportions of the tetragonal and orthorhombic
phases on cooling and on heating, are shown in Fig.~\ref{fig_07}.
The observed discontinuity in the unit cell volume at the transition
is $\sim1\%$. The $50:50$ proportions of the phases are reached at
$\rm\sim89.7\,K$ on cooling and at $\rm\sim107.2\,K$ on heating. The
width of the hysteresis is thus $\rm\sim17.5\,K$.

\subsection{\label{alternative}Possible alternative structure models}

Dealing with the neutron powder dataset with limited intensity does
not in principle exclude the possibility of having missed the
lowering of the symmetry just due to the fact that the very weak
diffraction intensities may potentially be confused with systematic
absences. Having carried out an exhaustive search in the
orthorhombic syngony for alternative crystal structure models of the
LT phase of \bcs, and upon carrying out careful refinements of their
parameters, we can as well list all other possible structure models
providing comparable agreement factors. It turns out that all
alternative possibilities which deserve being considered are limited
to the subgroups of the space group $Ibam$.

\begin{table}
\caption{\label{tab:table_3} List of all possible space groups for
the crystal structure of the LT phase of \bcs. SG and Set denote the
space group number and particular setting. $N_{a}$ and $N_{p}$ are
the total numbers of independent atoms and of refined positional
parameters (atomic coordinates). $N_{Cu}$ and $N_D$ denote the
numbers of independent $\rm Cu$ atoms and of the unique \cucu\ dimer
layers in each model correspondingly. \cucu\ distances for different
dimer layers are separated by slashes. The agreement factors
$\chi^2,R_p$ correspond to the refinements based on neutron powder
diffraction data collected at $\rm T=13\,K$.}
\begin{ruledtabular}
\begin{tabular}{@{}r@{ }l@{}|@{}c@{}|@{}c@{}|@{}c@{}|@{}c@{}}
SG&Set&$N_{a},N_{p}$&$N_{Cu},N_{D}$&\cucu,$\ang$&$\chi^2,R_p$\\
\hline
72&$Ibam$&$14, 30$&$2, 2$&$2.701(6)/2.774(6)$&2.66, 4.80\\
60&$Pbcn$ \footnotemark[1]&$21, 59$&$2, 2$&$2.713 (6)/2.768 (6)$&2.32, 4.41\\
57&$Pbcm$ \footnotemark[1]&$25, 63$&$2, 2$&$2.705 (6)/2.770 (6)$&2.55, 4.69\\
56&$Pccn$ \footnotemark[1]&$22, 58$&$4, 2$&$2.71 (2)/2.78 (3)$&2.42, 4.53\\[3pt]
\multirow{2}{*}{55}&\multirow{2}{*}{$Pbam$ \footnotemark[1]}&\multirow{2}{*}{$26, 62$}&\multirow{2}{*}{$4, 4$}&$2.71 (2)/2.70 (2)/$&\multirow{2}{*}{2.59, 4.75}\\
&&&&$2.86 (2)/2.70 (2)$&\\[3pt]
\multirow{2}{*}{50}&\multirow{2}{*}{$Pban$ \footnotemark[1]}&\multirow{2}{*}{$24, 56$}&\multirow{2}{*}{$4, 4$}&$2.71 (2)/2.77 (3)/$&\multirow{2}{*}{2.36, 4.45}\\
&&&&$2.70 (2)/2.77 (3)$&\\[3pt]
\multirow{2}{*}{49}&\multirow{2}{*}{$Pccm$ \footnotemark[1]}&\multirow{2}{*}{$28, 60$}&\multirow{2}{*}{$4, 2$\footnotemark[2]}&$2.73 (3) \:and\: 2.82 (2)/$&\multirow{2}{*}{2.46, 4.59}\\
&&&&$2.69 (3) \:and\: 2.72 (3)$&\\[3pt]
46&$I2cm$ \footnotemark[1]&$25, 62$&$2, 2$&$2.705 (6)/2.767 (6)$&2.50, 4.61\\
45&$Iba2$ \footnotemark[1]&$22, 57$&$4, 2$&$2.71 (3)/2.78 (3)$&2.36, 4.46\\
33&$P2_1cn$ \footnotemark[3]&$40, 119$&$4, 2$&$$2.71 (3)/2.77 (4)$$&2.26, 4.32\\[3pt]
\multirow{2}{*}{32}&\multirow{2}{*}{$Pba2$ \footnotemark[3]}&\multirow{2}{*}{$44, 115$}&\multirow{2}{*}{$8, 4$}&$2.78 (6)/2.76 (8)/$&\multirow{2}{*}{2.23, 4.31}\\
&&&&$2.67 (8)/2.80 (8)$&\\[3pt]
\multirow{2}{*}{30}&\multirow{2}{*}{$P2an$ \footnotemark[3]}&\multirow{2}{*}{$42, 117$}&\multirow{2}{*}{$4, 4$}&$2.73 (3)/2.77 (3)/$&\multirow{2}{*}{2.19, 4.23}\\
&&&&$2.68 (3)/2.77 (3)$&\\[3pt]
29&$Pca2_1$ \footnotemark[3]&$40, 119$&$4, 2$&$2.74 (6)/2.76 (6)$&2.12, 4.17\\[3pt]
\multirow{2}{*}{28}&\multirow{2}{*}{$P2cm$ \footnotemark[3]}&\multirow{2}{*}{$50, 125$}&\multirow{2}{*}{$4, 2$\footnotemark[2]}&$2.67 (4) \:and\: 2.74 (4)/$&\multirow{2}{*}{2.29, 4.34}\\
&&&&$2.76 (5) \:and\: 2.78 (5)$&\\[3pt]
\multirow{2}{*}{27}&\multirow{2}{*}{$Pcc2$ \footnotemark[3]}&\multirow{2}{*}{$44, 115$}&\multirow{2}{*}{$8, 2$\footnotemark[2]}&$2.76 (9) \:and\: 2.78 (9)/$&\multirow{2}{*}{2.14, 4.20}\\
&&&&$2.76 (6) \:and\: 2.69 (6)$&\\[3pt]
\multirow{2}{*}{26}&\multirow{2}{*}{$P2_1am$ \footnotemark[3]}&\multirow{2}{*}{$48, 127$}&\multirow{2}{*}{$4, 4$}&$2.69 (3)/2.70 (3)/$&\multirow{2}{*}{2.36, 4.42}\\
&&&&$2.86 (2)/2.70 (3)$&\\[3pt]
\multirow{2}{*}{23}&\multirow{2}{*}{$I222$ \footnotemark[1]}&\multirow{2}{*}{$24, 56$}&\multirow{2}{*}{$4, 2$\footnotemark[2]}&$2.70 (3) \:and\: 2.71 (3)/$&\multirow{2}{*}{2.43, 4.51}\\
&&&&$2.75 (3)$&\\[3pt]
\multirow{2}{*}{18}&\multirow{2}{*}{$P22_12_1$ \footnotemark[3]}&\multirow{2}{*}{$42, 118$}&\multirow{2}{*}{$4, 2$\footnotemark[2]}&$2.67 (3) \:and\: 2.74 (3)/$&\multirow{2}{*}{2.23, 4.28}\\
&&&&$2.77 (5)$&\\[3pt]
\multirow{4}{*}{16}&\multirow{4}{*}{$P222$}&\multirow{4}{*}{$48, 112$}&\multirow{4}{*}{$8, 4$\footnotemark[2]}&$2.85 (4) \:and\: 2.70 (4)/$&\multirow{4}{*}{2.22, 4.25}\\
&&&&$2.78 (6) \:and\: 2.75 (6)/$&\\
&&&&$2.68 (4) \:and\: 2.62 (5)/$&\\
&&&&$2.78 (6) \:and\: 2.75 (6)$&\\
\end{tabular}
\footnotetext[1]{Subgroup of space group 72 ($Ibam$).}
\footnotetext[2]{Different \cucu\ distances within one or more dimer
layers.} \footnotetext[3]{Subgroup of a few space groups which are
subgroups of space group 72 ($Ibam$).}
\end{ruledtabular}
\end{table}

In Table~\ref{tab:table_3}, we have summarized all hypothetically
possible models for the structure of the LT phase of \bcs\ providing
plausible refinements, along with the key values -- the numbers of
the independent unique \cucu\ dimer layers, the lengths of \cucu\
dimers, some statistical indicators for the complexities of the
models, and the corresponding agreement factors of the refinements.
We find it inappropriate to present here all the structure models in
details. The group-subgroup transformations are straightforward, and
any models will be supplied on request. The general trend is that
all alternative possibilities provide a pattern of the \cucu\ dimer
layers either very similar to or even more complex than that in our
model with the space group $Ibam$ (Tables~\ref{tab:table_1}
and~\ref{tab:table_2}). For some of them we have not 2 but 4
independent \cucu\ dimer layers. Since the apparent improvement in
agreement factors of the refinements for lower symmetries is only
achieved when the number of independent atoms and refinable
parameters is getting very high, and due to the obvious similarity
to the highest symmetry model with the space group $Ibam$
(Tables~\ref{tab:table_1} and~\ref{tab:table_2}), we strongly tend
to believe that the true structure of the LT phase of \bcs\ may
sufficiently well and adequately be described by our model with the
space group $Ibam$. Some later studies with higher precision may one
day show slight deviations from this model to one of those listed in
Table~\ref{tab:table_3}.

\section{\label{Summary}DISCUSSION AND SUMMARY}

We have investigated the low-temperature structural phase transition
in \bcs\ by means of powder synchrotron x-ray and neutron
diffraction and have determined the most probable average crystal
structure of its LT phase. The proposed crystal structure model of
the LT phase of \bcs\ has space group $Ibam$, and contains two
individual Cu atom positions (with only one in the room-temperature
structure). These two Cu atoms are forming two types of \cucu\
dimers thus altering the most essential magnetic \cucu\ exchange
path lengths in a stacking manner along the $c$ axis. The interlayer
\cucu\ exchange distances are also becoming barely different at low
temperature, forming in total two distinct but close interlayer
\cucu\ distances. In agreement with the previous study
(Ref.~\onlinecite{Samulon-prb-2006}), we confirm the phase
transition to be of first order, with a unit cell volume
discontinuity of $\sim1\%$ at the transition temperature, and a
temperature hysteresis of $\rm\sim17.5\,K$ (the $50:50$ phase weight
proportions being achieved at $\rm\sim89.7\,K$ on cooling and at
$\rm\sim107.2\,K$ on heating). In the LT crystal structure, the
layers with shorter \cucu\ dimer distances ($\sim2.701\ang$ at
$\rm13\,K$) preserve the mutual rotation of the $\rm CuO_4$
coordination squares with respect to each other in the \cucu\
dumbbells and the squarelike shapes of the $\rm(SiO_{3})_{4}$
molecules around them. This feature which is characteristic for the
room-temperature structure is lost in the layers with longer $\rm
Cu-Cu$ bonds ($\sim2.774\ang$ at $\rm13\,K$); the coordination $\rm
CuO_{4}$ squares in them unfold the mutual rotation and are aligned
identically, and the $\rm(SiO_{3})_{4}$ molecules get strongly
elongated along one of the $a-b$ diagonals, forming a checkerboard
pattern of their elongation directions. Thus, in the LT crystal
structure of \bcs, the two distinct adjacent \cso\ layers with
different \cucu\ dimer distances and even different internal
geometries are alternating in a sequential manner along the $c$
axis. This finding corroborates the interpretation of the
low-temperature INS data~\cite{RueggINS2007} in terms of existence
of two strongly inequivalent dimer layers in \bcs\ with possible
further complex modifications to one of them. The lower energy
excitations reported in~\cite{RueggINS2007} would correspond to the
$\rm Cu2-Cu2$ dimer layers with shorter \cucu\ distances, while the
higher energy multimode excitations correspond to the $\rm Cu1-Cu1$
dimer layers with longer \cucu\ distances. The incommensurate
modulation of the crystal structure observed before in different
studies did not hinder the structure determination, since as shown
in Ref.~\onlinecite{Samulon-prb-2006} the intensities of the
satellite reflections are $\sim10^{4}$ times weaker than that of the
fundamental reflections, and thus they did not disturb our powder
diffraction based \emph{average} structure determination.

Here we want to conclude with a remark about the possibility of
Dzyaloshinskii-Moriya (DM) interactions for \bcs. All three, HT, RT,
and LT, structures  of  \bcs\ are centrosymmetric. However, at RT
there is no inversion center in the middle of the \cucu\ dimers, as
correctly pointed out by  Ref.~\onlinecite{Samulon-prb-2006}. The
same holds for LT $\rm Cu2-Cu2$ dimer layers with shorter \cucu\
distances. In contrast to this, the LT $\rm Cu1-Cu1$ dimer layer
with longer \cucu\ distances has similarly to the HT $I4/mmm$
structure the inversion symmetry in the middle of the Cu dimer and
the intradimer DM coupling would be forbidden. In the real structure
of \bcs, the incommensurate modulation may well remove this local
symmetry and make DM interaction allowed again.

We note that the results presented here provide a structural model
for the compound which justifies the interpretation of previous INS
and NMR data by a number of groups~\cite{RueggINS2007,
Kraemer-PRB-2007} in terms of two inequivalent dimer layers in \bcs\
with further complex modifications to one of the layers that may
partially relieve frustration. Furthermore, the detailed structure
information provided here will be important input for the ongoing
\emph{ab initio} calculations of the exchange interactions in this
fascinating material.

\begin{acknowledgments}
This work is based on the experiments carried out at the Swiss Light
Source synchrotron and at the Swiss Spallation Neutron Source SINQ,
Paul Scherrer Institute, Villigen, Switzerland. We acknowledge
discussions with C.~Berthier, M.~Horvati\'c, S.~Kr\"{a}mer, E.~Joon,
J.~Mesot, C.~Batista, F.~Mila, A.~Tsirlin, F.~Duc, and K.~Sparta.
Our special thanks go to C.~R\"{u}egg for numerous inspiring
discussions and his careful reading of the manuscript. Work in
Tallinn was supported by the Estonian Ministry of Education and
Research under Grants No.~SF0690029s09 and No.~SF0690034s09, and
Estonian Science Foundation under Grants No.~ETF8198 and
No.~ETF8440.
\end{acknowledgments}

\bibliography{bacusi2o6-lt-str_v5-5}

\end{document}